\documentclass[aps,prl,reprint,superscriptaddress]{revtex4-1}

\usepackage{graphicx}
\usepackage{color,amsmath}
\usepackage[normalem]{ulem}
\usepackage{hyperref}
\usepackage{lipsum}

\newcommand{\Vg}{V_{\mathrm{G}}}
\newcommand{\Ig}{I_{\mathrm{G}}}
\newcommand{\Vsd}{V_{\mathrm{SD}}}
\newcommand{\Isd}{I_{\mathrm{SD}}}
\newcommand{\Ic}{I_{\mathrm{C}}}
\newcommand{\Ir}{I_{\mathrm{R}}}
\newcommand{\Bp}{B_{\perp}}
\newcommand{\Ici}{I_{\mathrm{SW}i}}
\newcommand{\oVg}{\overline{V}_{\mathrm{G}}}

\begin{document}
\title{A superconducting switch actuated by injection of high energy electrons}
\author{M.~F.~Ritter}
\affiliation{IBM Research Europe, S\"aumerstrasse 4, 8803 R\"uschlikon, Switzerland}

\author{A.~Fuhrer}
\email[email: ]{afu@ibm.zurich.com}
\affiliation{IBM Research Europe, S\"aumerstrasse 4, 8803 R\"uschlikon, Switzerland}

\author{D.~Z.~Haxell}
\affiliation{IBM Research Europe, S\"aumerstrasse 4, 8803 R\"uschlikon, Switzerland}

\author{S.~Hart}
\affiliation{IBM T. J. Watson Research Center, Yorktown Heights, NY, USA}

\author{P.~Gumann}
\affiliation{IBM T. J. Watson Research Center, Yorktown Heights, NY, USA}

\author{H.~Riel}
\affiliation{IBM Research Europe, S\"aumerstrasse 4, 8803 R\"uschlikon, Switzerland}

\author{F.~Nichele}
\email[email: ]{fni@ibm.zurich.com}
\affiliation{IBM Research Europe, S\"aumerstrasse 4, 8803 R\"uschlikon, Switzerland}

\date{\today}

\begin{abstract}
Fast cryogenic switches with ultra-low power dissipation are highly sought-after for control electronics of quantum computers, space applications and next generation logic circuits. However, existing high-frequency switches are often bulky, lossy or require large source-drain and gate currents for operation, making them unsuitable for many applications and difficult to interface to semiconducting devices. Here we present an electrically controlled superconducting switch based on a metallic nanowire. Transition from superconducting to resistive state is realized by tunneling of high-energy electrons from a gate contact through an insulating barrier. Operating gate currents are several orders of magnitude smaller than the nanowire critical source-drain current, effectively resulting in a voltage-controlled device. This superconducting switch is fast, self-resets from normal to superconducting state, and can operate in large magnetic fields, making it an ideal component for low-power cryogenic applications and quantum computing architectures.

\end{abstract}
\maketitle

\begin{figure*}
\includegraphics[width=2\columnwidth]{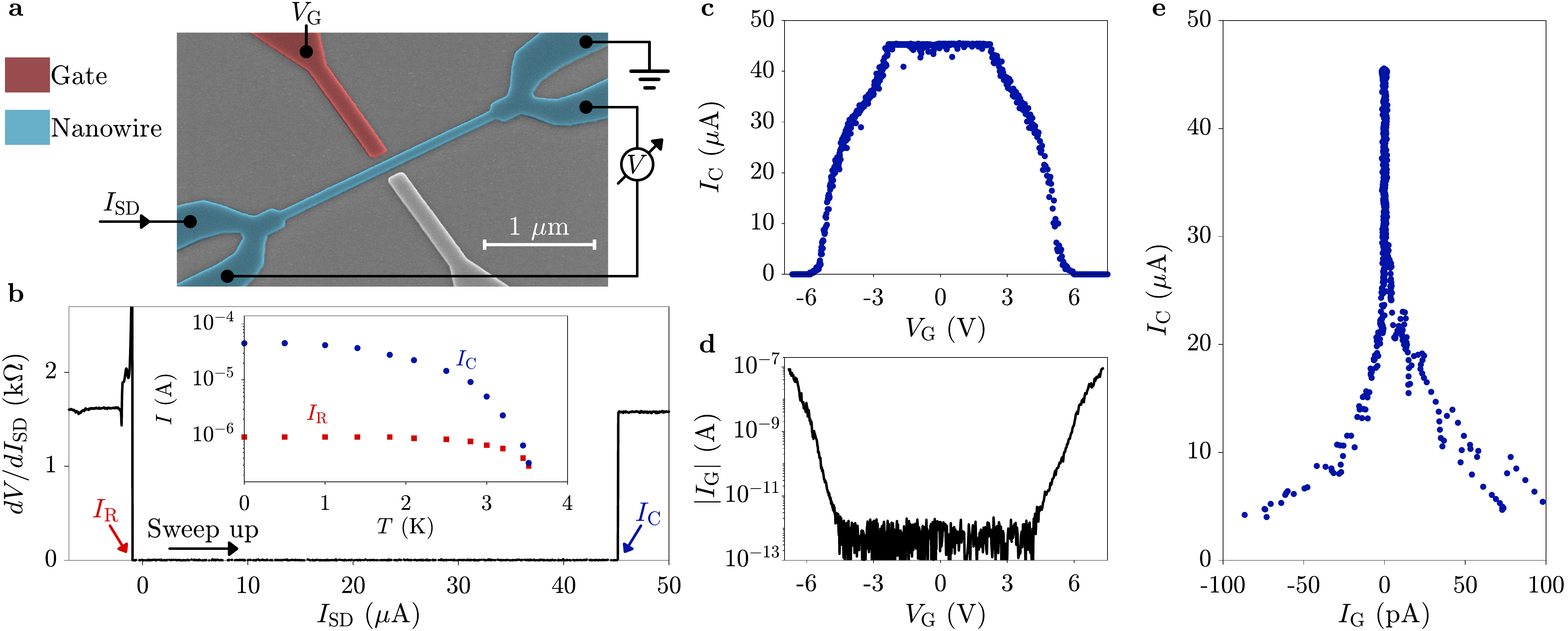}
\caption{\textbf{Operation of a metallic nanowire superconducting switch.}
\textbf{a} False-color scanning electron micrograph of a device identical to that under study, together with a schematic of the measurement setup. The silicon substrate is gray, the TiN nanowire blue and the gate electrode red. Another gate electrode (gray) was left grouded. \textbf{b} Differential resistance $dV/d\Isd$ of the nanowire as a function of $\Isd$, measured by sweeping up $\Isd$ starting from $-50~\mathrm{\mu A}$. Critical current $\Ic$ and retrapping current $\Ir$ are indicated. Inset: temperature dependence of $\Ic$ (blue dots) and $\Ir$ (red squares). \textbf{c} Critical current $\Ic$ in the nanowire as a function of gate voltage $\Vg$. \textbf{d} Absolute value of the gate current $\Ig$ flowing between gate and nanowire as a function of $\Vg$. A linear component $\Ig\sim 1~\mathrm{T\Omega}$, attributed to leakage in the measurement setup, was subtracted from the data (see Supplementary Information). \textbf{e} Parametric plot of $\Ic$ vs. $\Ig$. Obtained from the data in (c) and (d).}
\label{fig:1}
\end{figure*}

Superconducting circuits, thanks to their ultra-low power consumption and high speed, offer great promise as building blocks for quantum computing architectures and related cryogenic control electronics. In this context, it is especially intriguing to develop switching devices that can be electrically tuned between a superconducting and a resistive state at high frequency. Ultimately, such a three-terminal device would enable novel functionalities for which no semiconducting counterpart exists, such as cryogenic switches, ultra sensitive detectors, amplifiers, circulators, multiplexers and frequency tunable resonators~\cite{Faris1983,Pepe2000,Day2003,Lee2003,Leduc2010,HoEom2012,Hornibrook2015,Gasparinetti2015}. Several electrically controlled superconducting switches based on the injection of out-of-equilibrium quasiparticles in Josephson junctions have been realized~\cite{Wong1976,Morpurgo1998,Giazotto2005,Nevirkovets2009}. However, Josephson junctions typically come with limited source-drain critical currents and the requirement to operate in magnetic field free environments. Consequently, architectures which do not rely on Josephson junctions are subject to intense study. Such pioneering approaches are based on three or four terminal devices where electrical currents~\cite{Wagner2020}, locally generated Oersted fields~\cite{Buck1956} or heat~\cite{McCaughan2014,Toomey2019,McCaughan2019} drive a superconducting channel normal. Devices mentioned above, either based on Josephson junctions or alternative designs, are typically characterized by low gate impedance, so that large gate currents are needed for operation and substantial power is drawn from the driving circuit. Furthermore, these devices typically operate with large source-drain bias currents, which inhibit self-resetting from normal to superconducting state due to Joule heating. Switching then requires additional circuitry that periodically drives the source-drain current below the retrapping current, where a transition to the superconducting state occurs. When full suppression of the source-drain critical current is achieved, operation at low current bias is enabled and self-resetting from the normal to the superconducting state becomes possible, as recently shown with a novel electrothermal device~\cite{McCaughan2019}. Finally, recent experiments suggest that moderate electric fields might affect superconductivity in metallic nanowires~\cite{DeSimoni2018,Paolucci2020}. As detailed below, our findings point to a different physical origin for the data of Refs.~\cite{DeSimoni2018,Paolucci2020}. 

Here we demonstrate a superconducting switch with ultra-high gate impedance (much greater than $1~\mathrm{G\Omega}$), which operates via injection of high energy electrons through an insulating barrier and is capable of fast self-resetting. A typical device is shown in Fig~\ref{fig:1}(a): it consists of a metallic nanowire (blue) with a side gate (red), all deposited on an insulating substrate (gray). Application of a gate voltage $\Vg$ results in the flow of a current $\Ig$ between gate and nanowire, whose dependence is exponential on $\Vg$. Concomitant to the onset of $\Ig$, $\Ic$ in the nanowire is reduced and eventually suppressed. Detailed measurements indicate that relatively few electrons, injected at energies much higher than the superconducting gap, trigger the generation of a large number of quasiparticles and quench superconductivity. We found this mechanism to be extremely efficient, with suppression of the nanowire source-drain critical current $\Ic$ by tens of $\mathrm{\mu A}$ for gate currents below $1~\mathrm{pA}$. Total suppression of superconductivity is achieved with gate currents between tens of pA to a few nA (depending on the superconductor in use). This is in stark contrast to previously demonstrated devices~\cite{Morpurgo1998,McCaughan2014,McCaughan2019}, where gate currents comparable to the nanowire critical current (a few $\mathrm{\mu A}$) were needed for operation. In addition to these remarkable properties, devices presented here operate similarly for both positive and negative gate voltages, and for the entire temperature and magnetic field ranges in which superconductivity in the nanowire persists, making them applicable to a broader range of environments than Josephson junctions. Finally, our superconducting switch is extremely compact, fabricated in a single lithographic step, and is readily integrated in existing architectures, either directly on quantum processors as a signal router, or as a cryogenic interface between classical, voltage-driven, digital electronics and superconducting, current-driven circuits.
In this Article we introduce the basic device operation, in terms of critical current and its dependence on gate voltage, temperature and magnetic field. This basic characterization is performed with different substrates and superconductors. We then investigate how injected quasiparticles influence superconductivity along the length of the channel, and demonstrate fast electrical switching and self-resetting to the superconducting state. After presenting the experimental observations, we will elaborate on their physical origins and their implications for device designs.

Figure~\ref{fig:1}(a) shows a typical device, together with the measurement setup in use. The device consists of a $2~\mathrm{\mu m}$ long, $80~\mathrm{nm}$ wide TiN wire (blue) with a TiN side gate (red). Wire and side gate are separated by an $80~\mathrm{nm}$ wide gap. Gates and nanowires were defined by electron beam lithography and dry etching of a TiN film deposited on an intrinsic Si substrate. Measurements were performed by low-frequency lock-in techniques by passing a source-drain current $\Isd$ in the nanowire and recording the resulting voltage $V$. Gate voltage $\Vg$ was applied via a source-measure unit, which also recorded the current $\Ig$ entering the gate contact. A second side gate (gray) was not operated in this work and left grounded. Further details on materials, samples fabrication and measurement techniques are reported in the Methods Section.
At $\Vg=0$, the nanowire showed a critical current $\Ic=45~\mathrm{\mu A}$, measured sweeping $\Isd$ in either direction starting in the superconducting state. In contrast, when sweeping $\Isd$ from the normal state towards zero, superconductivity was re-established below the retrapping current $\Ir=1~\mathrm{\mu A}$. Figure~\ref{fig:1}(b) shows the nanowire differential resistance $dV/d\Isd$, measured while sweeping $\Isd$ in the positive direction. The inset gives the temperature dependence of $\Ic$ and $\Ir$. The large difference between $\Ic$ and $\Ir$, especially marked at low temperature, is likely due to self-heating when the nanowire is in the normal state, together with the difficulty in extracting heat via the substrate or the leads~\cite{Li2011}. Figures~\ref{fig:1}(c) and (d) show $\Ic$ and $\Ig$, respectively, as a function of $\Vg$. For $\Vg\sim\pm2.5~\mathrm{V}$, just before $\Ig$ reached detection level ($\sim100~\mathrm{fA}$ in our setup), $\Ic$ started decreasing. At $\Vg=\pm5.5~\mathrm{V}$, where $\Ig\approx\pm 1.5~\mathrm{nA}$, $\Ic$ vanished and the nanowire reached its normal state resistance of $1.6~\mathrm{k\Omega}$. The parametric plot of $\Ic$ vs. $\Ig$ shown in Fig.~\ref{fig:1}(e) indicates an extremely sharp suppression of $\Ic$ for small values of $\Ig$ (about $50\%$ of the suppression took place within the noise level for $\Ig$), followed by a slower decay which persisted up to $\Ig\sim1.5~\mathrm{nA}$.

\begin{figure}
\includegraphics[width=\columnwidth]{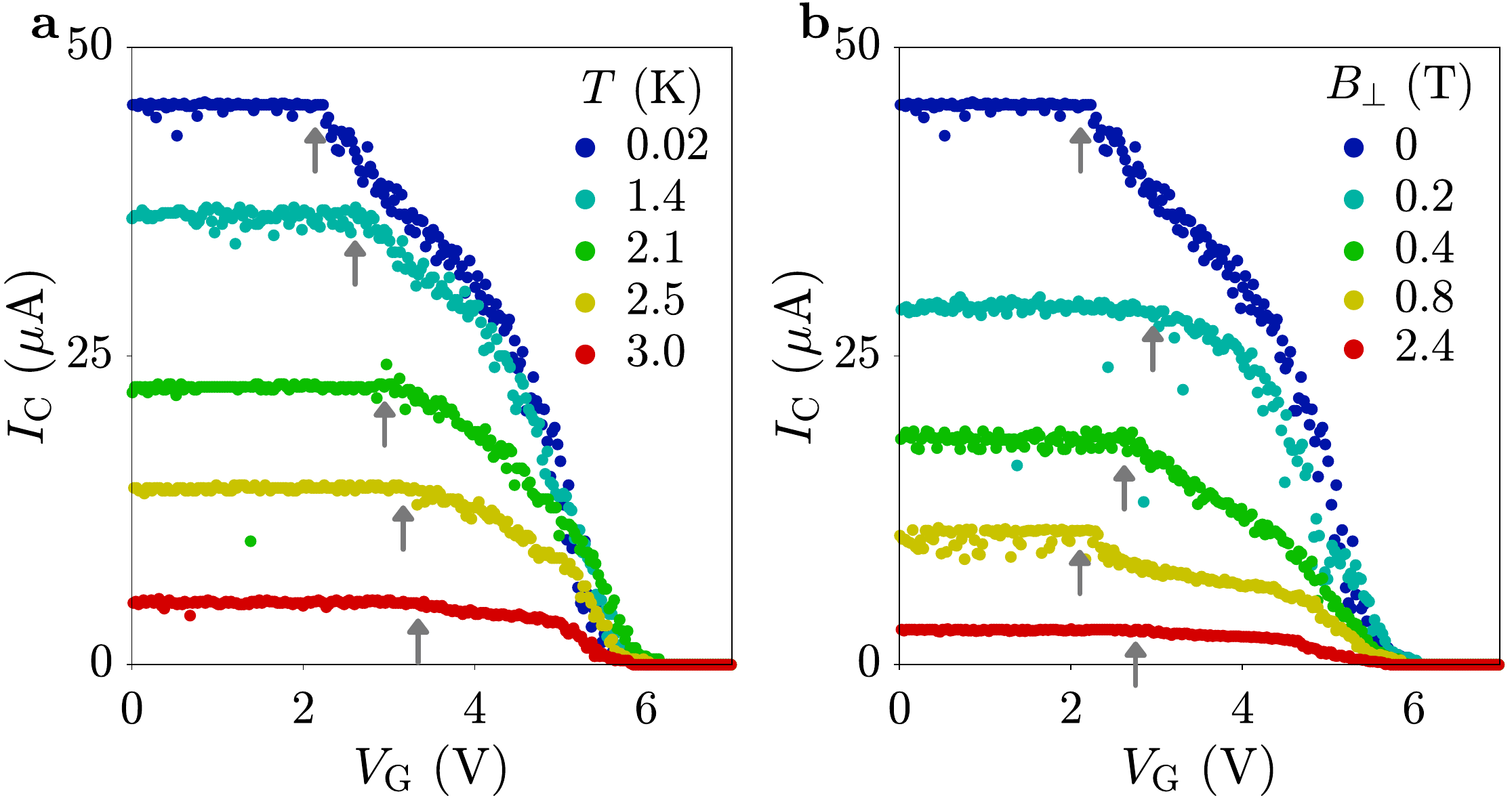}
\caption{
\textbf{Temperature and magnetic field dependence.}
\textbf{a} Critical current $\Ic$ of the device presented in Fig.~\ref{fig:1} as a function of gate voltage $\Vg$ for various temperatures $T$ (see legend). \textbf{b} Critical current $\Ic$ of the device presented in Fig.~\ref{fig:1} as a function of out-of-plane magnetic fields $\Bp$ (see legend). Gray arrows indicate the gate voltage where $\Ic$ starts decreasing.}
\label{fig:2}
\end{figure}

Figure~\ref{fig:2}(a) and (b) show the gate voltage dependence of $\Ic$ for various temperatures $T$ and out-of-plane magnetic fields $\Bp$, respectively. Neither temperature nor field affected the $\Ig$ vs. $\Vg$ characteristics of Fig.~\ref{fig:1}(d) (see Supplementary Information), and resulted in identical $\Vg$ values for complete suppression of superconductivity in the nanowires, up to the critical temperature and critical field. On the other hand, the onset of $\Ic$ suppression systematically moved to higher $\Vg$ for higher temperatures (see gray arrows). A more complicated dependence was observed as a function of $\Bp$.

\begin{figure}
\includegraphics[width=\columnwidth]{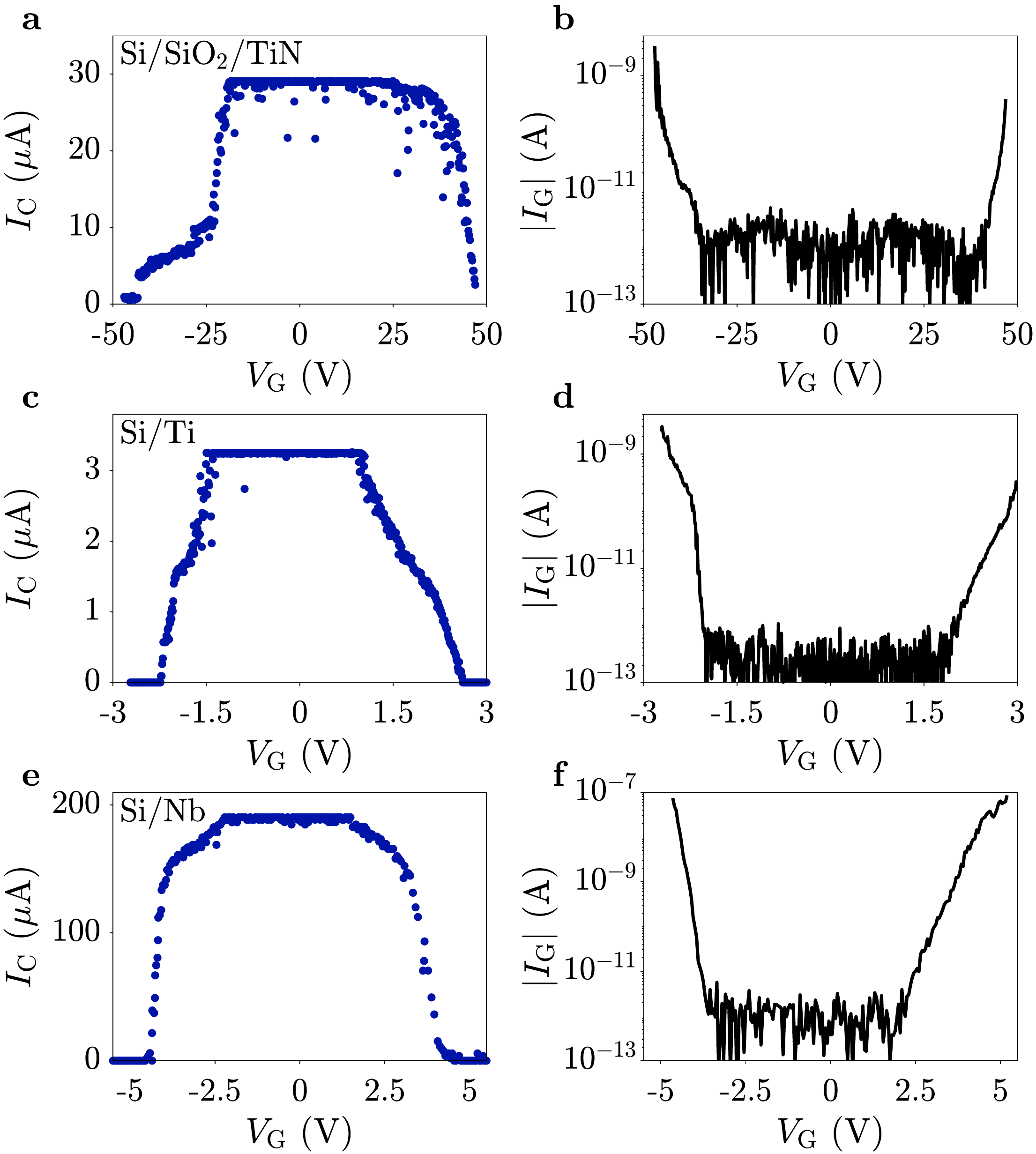}
\caption{\textbf{Critical current suppression in various superconductors.}
\textbf{a,b} Critical current $\Ic$ and gate current $\Ig$ as a function of gate voltage $\Vg$ for a TiN wire on a $25~\mathrm{nm}$ thick $\mathrm{SiO_2}$ film thermally grown on Si substrate. The wire is $2~\mathrm{\mu m}$ long, $80~\mathrm{nm}$ wide and $20~\mathrm{nm}$ thick. \textbf{c,d} Critical current $\Ic$ and gate current $\Ig$ as a function of gate voltage $\Vg$ for a Ti wire on Si substrate. The wire is $2~\mathrm{\mu m}$ long, $200~\mathrm{nm}$ wide and $30~\mathrm{nm}$ thick. \textbf{e,f} Critical current $\Ic$ and gate current $\Ig$ as a function of gate voltage $\Vg$ for a Nb wire on Si substrate. The wire is $2~\mathrm{\mu m}$ long, $200~\mathrm{nm}$ wide and $13~\mathrm{nm}$ thick.}
\label{fig:3}
\end{figure}

Suppression of $\Ic$ concomitant to, or slightly anticipating, the onset of $\Ig$ above detection level was confirmed for over 20 TiN devices, characterized by various gate shapes, nanowire widths ($40$, $80$ and $200~\mathrm{nm}$), nanowire lengths ($650~\mathrm{nm}$, $1$ and $2~\mathrm{\mu m}$), and gate-to-wire separations ($80$ and $160~\mathrm{nm}$). Some of these measurements are presented in the Supplementary Information. Similar behavior was also observed on devices with a different substrate than Si or with a different superconductor than TiN.
Figures~\ref{fig:3}(a) and (b) show measurements performed on a TiN device as that of Fig.~\ref{fig:1}(a), but deposited on a $25~\mathrm{nm}~\mathrm{SiO_2}$ layer thermally grown on Si. Despite the vastly different operational range of $\Vg$ with respect to that of Fig.~\ref{fig:1}, suppression of $\Ic$ still coincided with the onset of $\Ig$. Devices with a $\mathrm{SiO_2}$ interlayer further showed a characteristic asymmetry of the $\Ic$ vs. $\Vg$ curve, with a sharper suppression of $\Ic$ for negative than for positive $\Vg$. Given the sharp termination of the gate electrode, and the large electric field reached on $\mathrm{SiO_2}$ substrates, extraction of electrons from the gate is expected to be promoted for negative gate biases. In the present case, detection of small gate current asymmetries is hindered by spurious current leakage in the measurement setup for high gate biases. Figures~\ref{fig:3}(c) and (d) show $\Ic$ and $\Ig$, respectively, as a function of $\Vg$ for a Ti nanowire as that of Fig.~\ref{fig:1}(a), but with $200~\mathrm{nm}$ width and $30~\mathrm{nm}$ thickness. In this case, the normal state was reached for $\Ig$ as low as $30~\mathrm{pA}$ for positive $\Vg$. Figures~\ref{fig:3}(e) and (f) show $\Ic$ and $\Ig$, respectively, as a function of $\Vg$ for a Nb nanowire as that of Fig.~\ref{fig:1}(a) but with $13~\mathrm{nm}$ thickness. Similarly to the previous cases, $\Ic$ started decreasing with $\Ig$ still below $100~\mathrm{fA}$. However complete suppression of $\Ic$ required $\Ig\geq20~\mathrm{nA}$. Overall, these results indicate that the switching mechanism presented here is generic, and not linked to specific superconductors or substrates. On the other hand, data also suggests that small gap superconductors (e.g. Ti) require considerably less gate current for switching to occur with respect to superconductors with larger gaps (e.g. TiN or Nb).

\begin{figure*}
\includegraphics[width=2\columnwidth]{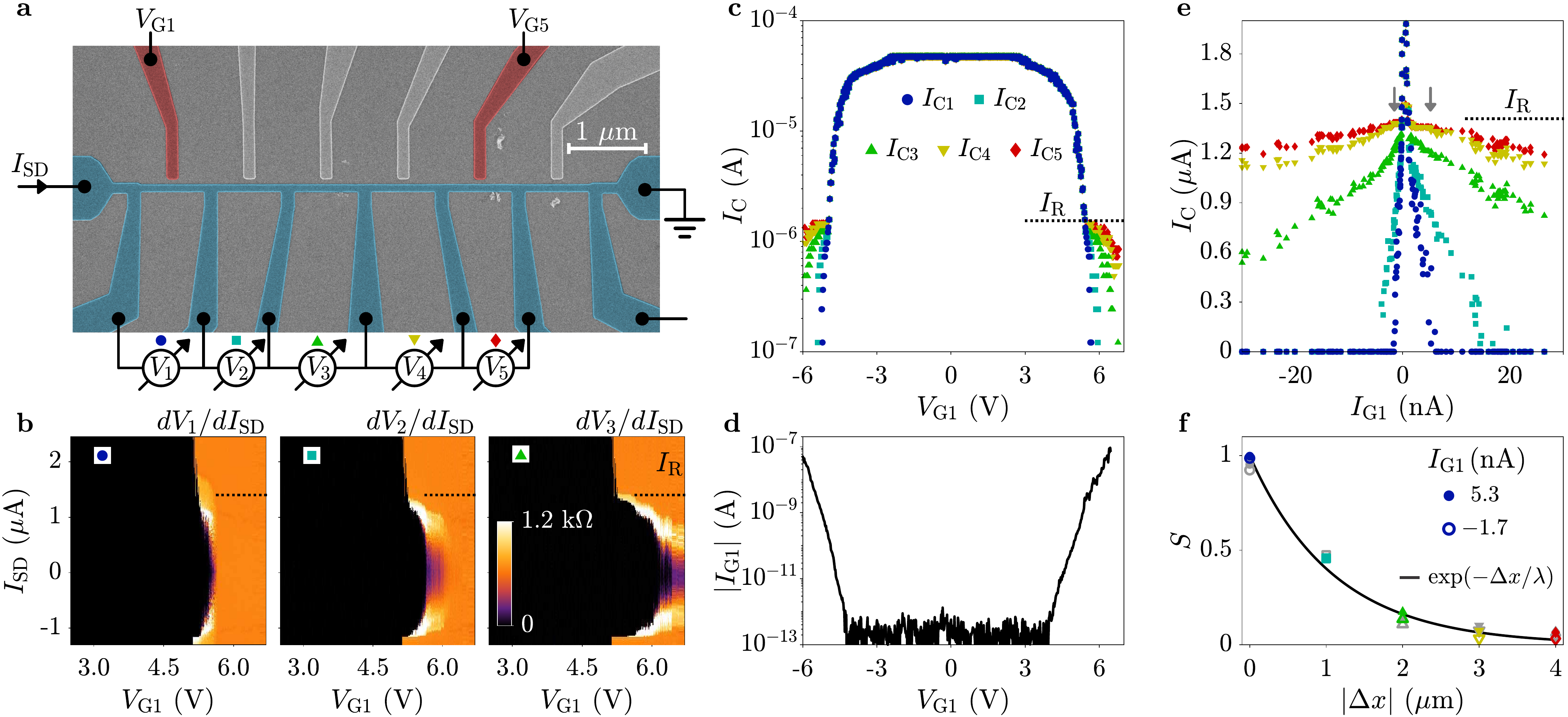}
\caption{\textbf{Spatially resolved suppression of the crtical current.}
\textbf{a} False-color scanning electron micrograph of a device as that under study, together with a schematic of the measurement setup. Colors are as in Fig.~\ref{fig:1}(a). Gray gates were grounded. \textbf{b} Differential resistances $dV_i/d\Isd$ for $i=1,2,3$ close to the complete suppression the superconducting state. The retrapping current $\Ir$ is indicated by a dotted line. \textbf{c} Critical currents $\Ici$ for $i=1,2,3,4,5$ as a function of gate voltage $V_{\mathrm{G1}}$. \textbf{d} Absolute value of gate current $I_{\mathrm{G1}}$ as a function of gate voltage $V_{\mathrm{G1}}$. \textbf{e} Parametric plot of $\Ici$ as a function of $I_{\mathrm{G1}}$. Gray arrows indicate the $I_{\mathrm{G1}}$ values where the suppression factor is calculated. \textbf{f} Suppression factor $S$ as a function of the distance between gate and nanowire segment. Colorful markers were obtained by sweeping $I_{\mathrm{G1}}$, gray markers by sweeping $I_{\mathrm{G5}}$. Full and empty markers refer to positive and negative gate polarities, respectively. The solid line is a fit to an exponential.}
\label{fig:4}
\end{figure*}

Measurements presented so far were conducted in relatively short nanowires, where sharp transitions from zero to the normal state were observed. In the following, we present how superconductivity is reduced from the electron injection point, along the nanowire length. The device shown in Fig.~\ref{fig:4}(a) consists of six TiN segments of $1~\mathrm{\mu m}$ length and $80~\mathrm{nm}$ width. A current $\Isd$ was injected along the nanowire, and the voltages $V_i$ across the first five segments were simultaneously recorded (the first side contact to the right was left floating). Figure~\ref{fig:4}(b) shows zoom-ins of the differential resistance of the first three segments for $\Isd<\Ir$. In addition to the varying extension of the zero-resistance state, segments far away from the injection point showed an intermediate region where the resistance is above zero, but below the normal state value. This intermediate regime indicates normal and superconducting parts coexisted close to each other in the same segment. Critical currents, with markers as indicated in Fig.~\ref{fig:4}(a), are shown in Fig.~\ref{fig:4}(c) as a function of $V_{\mathrm{G1}}$, with the corresponding gate current $I_{\mathrm{G1}}$ shown in Fig.~\ref{fig:4}(d). The parametric plot of $\Ic$ vs. $I_{\mathrm{G1}}$ is shown in Fig.~\ref{fig:4}(e), focusing on the region below the retrapping current $\Ir=1.4~\mathrm{\mu A}$.
Non-local measurements highlight two distinct regimes. For $\Isd>\Ir$, switching in all the segments was simultaneous: once a local hotspot was created, Joule heating warmed up the surrounding TiN in a runaway fashion and the entire channel turned normal. For $\Isd<\Ir$ sequential switching occurred: the closer a segment was to the biased gate, the less gate current was needed to reach the normal state.

In Fig.~\ref{fig:4}(f) we plot the critical current suppression factor, defined as $S=(\Ir-\Ic)/\Ic$, as a function of distance $|\Delta x|$ (measured between the point of injection and the center of each segment). Data are extracted for $I_{\mathrm{G1}}$ as shown in the legend (colorful markers), corresponding to $|\Ig|$ values where $I_{\mathrm{C1}}$ vanished (see vertical arrows in Fig.~\ref{fig:4}(e)). A similar analysis performed by sweeping $V_{\mathrm{G5}}$ is plotted in Fig.~\ref{fig:4}(f) with gray markers. Results are highly consistent for both gates and voltage polarities, and do not show dependence on the relative position between injection point, measurement point and ground contact. The limited data range and the relatively large point-to-point scattering does not allow us to determine the functional form of $S(\Delta x)$. A fit to an exponentially decaying function $\exp{(-\Delta x/\lambda)}$ (solid line in Fig.~\ref{fig:4}(f)), yields a characteristic decay length $\lambda\sim1.2~\mathrm{\mu m}$.

\begin{figure}
\includegraphics[width=\columnwidth]{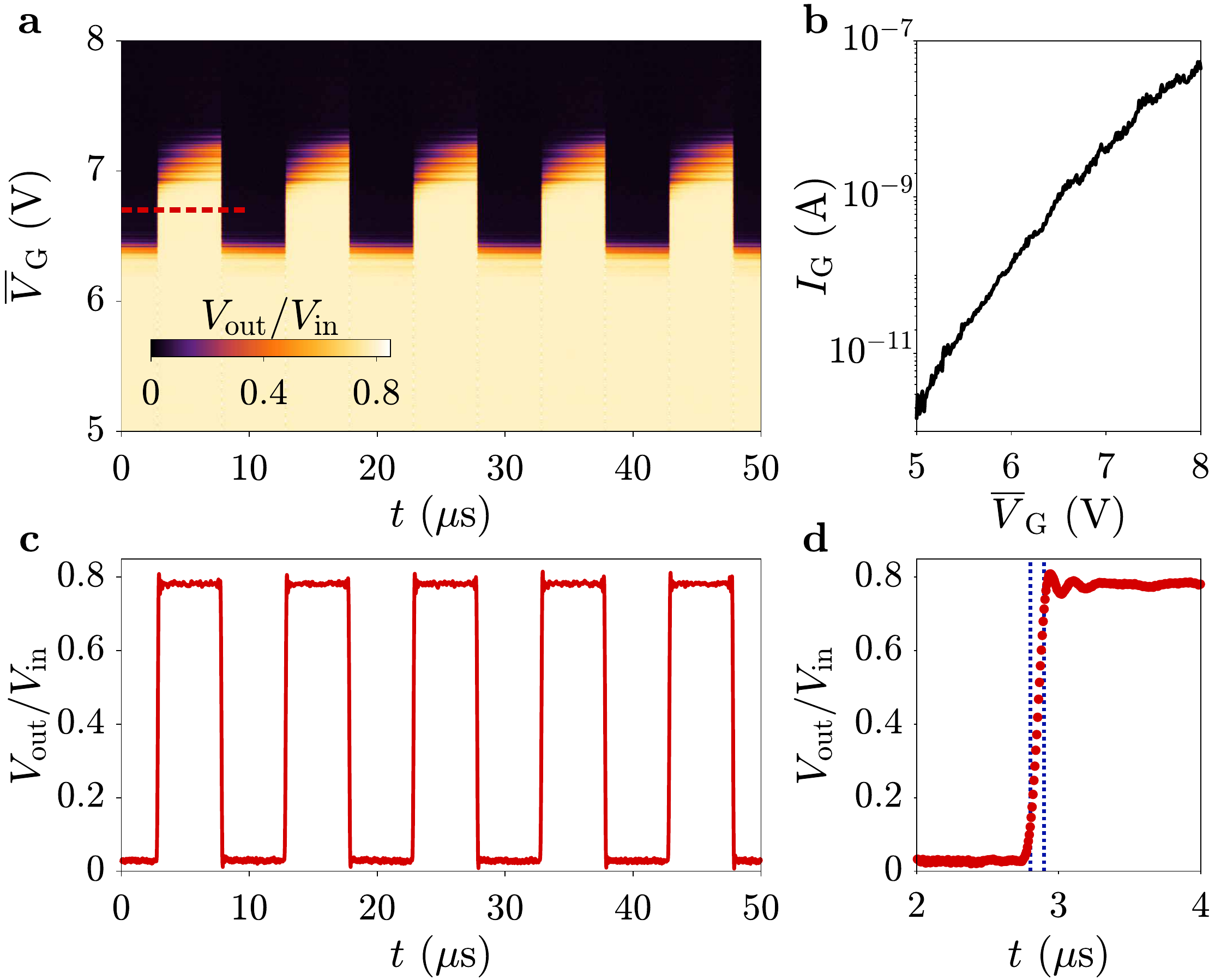}
\caption{\textbf{Fast switching in a metallic nanowire.}
\textbf{a} Time-dependent switching characteristics of a device as that of Fig.~\ref{fig:1}(a) as a function of DC gate voltage $\oVg$. A $100~\mathrm{kHz}$ square wave with peak-to-peak amplitude of $1~\mathrm{V}$ was added to $\oVg$. \textbf{b} Gate current $\Ig$ as a function of $\oVg$ measured simultaneously to the data in (a). \textbf{c} Line-cut of the data in (a) for $\oVg=6.7~\mathrm{V}$ (see red line). \textbf{d} Zoom-in of the data in (c) in proximity to a normal to superconducting state transition. Dashed vertical lines indicate the $10\%$ to $90\%$ amplitude transition, corresponding to $90~\mathrm{ns}$. This value is limited by the measurement bandwidth of the setup in use and serves as upper limit for the device switching time.}
\label{fig:5}
\end{figure}

The switching speed of a device similar to that of Fig.~\ref{fig:1}(a) was tested by recording the transmission of a $250~\mathrm{MHz}$ sinusoidal signal across the nanowire as a function of time. The nanowire resistance was modulated by adding a square wave with peak-to-peak amplitude of $1~\mathrm{V}$ and repetition rate of $100~\mathrm{kHz}$ to the DC gate voltage $\oVg$ (see the Methods section for further detail and the Supplementary Information for a circuit diagram). The ratio between the transmitted voltage $V_{\mathrm{out}}$ and the voltage input to the measurement setup $V_{\mathrm{in}}$ is shown in Fig.~\ref{fig:5}(a) as a function of time $t$ and $\oVg$, with the time-averaged gate current shown in Fig.~\ref{fig:5}(b). Clear switching operation was achieved within a $500~\mathrm{mV}$ interval around $\oVg=6.7~\mathrm{V}$, corresponding to a gate current of $1~\mathrm{nA}$. Figure~\ref{fig:5}(c) shows a line cut of Fig.~\ref{fig:5}(a) for $\oVg=6.7~\mathrm{V}$, demonstrating fast and reproducible switching between two impedance states. A zoom-in close to a rise point is shown in Fig.~\ref{fig:5}(d), with dashed lines marking the transition between $10\%$ and $90\%$ of the step height, which takes place in $90~\mathrm{ns}$ (similar results are obtained for the decay time). Such transient equals three times the time constant of the lock-in amplifier used for these measurements ($30~\mathrm{ns}$) and is taken as the shortest switching time measurable with the setup in use, and as the upper limit for the device response time. Future work will take advantage of samples specifically designed for microwave measurements~\cite{Wagner2020} and correlation techniques~\cite{McCaughan2014} to test the ultimate switching speed of the device.

After presenting the device operation, we now discuss possible origins for the observed phenomena. Electrons emitted from the gate reach the nanowire in a deeply out-of-equilibrium state, with energies of the order $e\Vg$, much larger than the superconducting gap ($500~\mathrm{\mu eV}$ for TiN~\cite{Pracht2012}). As electrons relax to the gap edge by inelastic scattering with other electrons and phonons, a large number of quasiparticles are generated within the nanowire. A sufficiently high concentration of quasiparticles drives the nanowire normal by quenching the superconducting gap~\cite{Semenov2001} and suppression of the depairing critical current~\cite{Semenov2005}, leading to a switch to the normal state. Such behavior is reminiscent of superconducting nanowire single photon detectors (SNSPDs)~\cite{Kadin1996,Goltsman2001}, where the strike of a visible or infrared photon promotes single electrons to high energies, which in turn trigger the generation of a large amount of quasiparticles as they relax. In the present case, high energy electrons are provided directly by the gate current, without the need of photons. We showed that the critical current suppression is highest at the point of injection and decreases with distance, with a characteristic length scale of the order $1~\mathrm{\mu m}$. This length scale is presumably related to the diffusion length of high energy electrons and the relaxation length of generated quasiparticles. A framework for calculating quasiparticle density profiles has been put forward for SNSPDs in Ref.~\onlinecite{Engel2013}.

Other mechanisms are ruled out as source of the suppression of $\Ic$. Joule heating in the barrier due to dissipated power $\Vg\Ig$ and consequent local temperature increase is at odds with the absence of temperature dependence for full suppression of $\Ic$ in Fig.~\ref{fig:2}(a). Recent work argued on the effect of electric fields on the critical current of metallic nanowires, using a similar device as that of Fig.~\ref{fig:1}(a)~\cite{DeSimoni2018,Paolucci2020}.
Our results resemble data in Refs.~\onlinecite{DeSimoni2018,Paolucci2020}, including ambipolar behavior, response to temperature and magnetic field and spacial suppression of the the supercurrent reduction. However, electric field induced suppression of superconductivity is excluded, in our work, due to the very small gate voltages applied, the consistent correlation between gate currents and critical currents, and the pronounced non-local responses which extend far beyond the gate induced electric field (see Fig.~\ref{fig:4}). We note that electric field modulation of superconductivity has been previously demonstrated in metallic thin films~\cite{Glover1960,Choi2014,Piatti2017}. However, changes of critical temperature by less than one percent required significantly stronger electric fields than those applied here. On the other hand, we showed that gate currents of a few pA or less lead to a strong suppression of the critical current of metallic nanowires. This effect is especially strong for small gap superconductors on highly insulating substrates, as investigated in Refs.~\onlinecite{DeSimoni2018,Paolucci2020}. For large gap superconductors on semi-insulating substrates, i.e. the main focus of this work, the lower energy of emitted electrons, together with the requirement of more quasiparticle density to suppress superconductivity, make required gate currents larger and their detection more feasible. 

Our findings open the door to fast electrical switches with a resistive and a superconducting state, which operate at extremely low gate currents. The ultra-high gate impedance and low power consumption makes our switches ideal as an interface between voltage driven transistors and current driven superconducting circuits. The quasiparticle relaxation length limits the extent of the segment that can be switched, and consequently the largest normal state resistance of the device. In this context, choosing superconductors with large diffusion length and large normal state resistivity is desirable. Alternatively, nanowires of arbitrary length can be operated by choosing $\Isd>\Ir$~\cite{McCaughan2014,Wagner2020}. In this situation, switching from the superconducting to the normal state requires actively resetting $\Isd$ to values smaller than $\Ir$. For low current bias, a response time below $90~\mathrm{ns}$ was measured, limited by the setup in use. The ultimate switching speed is likely determined by the quasiparticle relaxation time in the superconductor, typically below $1~\mathrm{ns}$~\cite{Ilin1998,Annunziata2010,Beck2011}. Furthermore, the ability to adjust the quasiparticle density in the superconducting state with the gate current enables tuning of the kinetic inductance of nanowire resonators~\cite{Leduc2010}. This paves the way to applications of such tunable resonators as coupling elements or filters in a superconducting qubit platform.
Devices studied here could also serve as tools for novel studies of quasiparticle physics at unprecedented energy scales and in the limit of no current flowing in the nanowire. Temperature and magnetic field studies shown in Fig.~\ref{fig:2} indicate that the gate current needed to completely suppress $\Ic$ is largely unaffected by closure of the superconducting gap, which is desirable for device applications. However, Figs.~\ref{fig:2}(a) and (b) show richer physics at lower values of $\Ig$, with the initial suppression of $\Ic$ moving to higher and higher gate voltages as temperature increases (see gray arrows). This behavior reflects the increase of quasiparticle density in the wire with temperature, requiring more electrons to be injected before a sizeable effect on $\Ic$ is observed. Systematic studies of the more complicated variations of $\Ic$ vs. $\Ig$ as a function of $\Bp$ could shed new light on the physics of field repulsion and vortex penetration in nanowires~\cite{Engel2013}. Furthermore, combining length dependence studies as in Fig.~\ref{fig:3} with time-resolved experiments as in Fig.~\ref{fig:5}, will provide a novel tool to investigate quasiparticle dynamics and thermal effects.

In conclusion, we presented a metallic nanowire switch where superconductivity is quenched by gate currents several orders of magnitude smaller than the source-drain critical current. Devices could be operated in non-latching mode and on fast time scales. Due to the generality and robustness of the presented effect, and the easy device fabrication, our findings can be put to immediate use in quantum computing architectures and cryogenic electronics. 

\textbf{Acknowledgments}
We thank A.~Pushp, B.~Madon and M.A.~Mueed for deposition of the TiN films. We thank G.~Salis and W.~Riess for fruitful discussions, A.~Olziersky, S.~Paredes and U.~Drechsler for technical help. F.~Nichele acknowledges support from the European Research Commission, grant number 804273.

\textbf{Author contributions}
F.N. conceived the experiments. A.F. and F.N. designed the samples. S.H and P.G. deposited the Ti and Nb films. M.F.R. and D.Z.H. fabricated the devices. M.F.R., A.F., D.Z.H. and F.N. performed the measurements. M.F.R., A.F., D.Z.H, H.R. and F.N. interpreted and analyzed of the data. A.F. and F.N. wrote the manuscript.

\section*{Methods}
\textbf{Sample fabrication}
Devices were obtained on either intrinsic Si substrates or intrinsic Si substrates with a $25~\mathrm{nm}$ thermally grown $\mathrm{SiO_2}$ top layer. Both high and low resistivity Si resulted in similar device performance at low temperature in terms of gate currents vs. gate voltages. Just prior to the deposition of the superconductor, Si chips were immersed in a buffered HF solution for removing the native Si oxide. Nanowires were defined by electron beam lithography, as detailed below. After the nanowires were fabricated, Ti/Au bonding pads, placed about $170~\mathrm{\mu m}$ away from the active region of the devices, were defined by optical lithography, thermal evaporation and lift-off.
\textit{TiN wires}: a $20~\mathrm{nm}$ TiN film was deposited by DC reactive magnetron sputtering. A $50~\mathrm{nm}$ thick layer of hydrogen silsesquioxane (HSQ) based negative tone resist was used as mask. After resist development, unprotected TiN areas were removed by inductively coupled plasma etching in HBr plasma. After etching of the TiN, HSQ was removed by immersion in a diluted hydrofluoric acid solution. Characterization of the TiN film gave a resistivity of $68~\mathrm{\Omega}$ per square, a critical temperature of $3.7~\mathrm{K}$ and a critical out-of-plane magnetic field of $3.5~\mathrm{T}$. These properties remained unchanged in the completed devices.
\textit{Ti wires}: Ti wires were defined by electron beam lithography on a positive tone polymethyl methacrylate mask, electron beam evaporation of a $30~\mathrm{nm}$ thick Ti film and lift-off. Ti evaporation was performed at a base pressure of $10^{-9}~\mathrm{mbar}$ and at a deposition rate of $1~\mathrm{nm~s^{-1}}$. The high deposition rate was chosen to minimize contamination of the Ti film during evaporation~\cite{Wang2018}. The Ti wire of Fig.~\ref{fig:3}(c) had a critical temperature of $220~\mathrm{mK}$ and a normal state resistance of $74~\mathrm{\Omega}$.
\textit{Nb wires}: Nb wires were obtained by sputtering of a $13.5~\mathrm{nm}$ film on intrinsic Si substrates and following a fabrication similar to that described above for TiN. Dry etching was performed with $\mathrm{Ar/Cl_2}$ plasma. The Nb wire of Fig.~\ref{fig:3}(e) had a normal state resistance of $655~\mathrm{\Omega}$.

\textbf{Electrical measurements}
Electrical measurements were performed in a dilution refrigerator with base temperature of $10~\mathrm{mK}$. Low-pass RC filters and microwave pi-filters were installed along each line. A DC source-drain current $\Isd$, superimposed to a small AC component of $30~\mathrm{nA}$ and $113~\mathrm{Hz}$ was applied to the nanowire via large bias resistors. The AC differential voltage $V$ across the nanowire was then recorded with lock-in amplifiers with $10~\mathrm{M\Omega}$ input impedance and used to calculate the differential resistance $dV/d\Isd$. Measurements were recorded with $\Isd$ as the fast axis, sweeping from zero to positive values. This allowed initializing the wires to the superconducting state before each sweep started. Gate voltages were applied with a Keysight B2902A source-measure unit, which also recorded the current entering the gate contact. To avoid damaging the devices, a compliance of $\pm95~\mathrm{nA}$ was chosen. A linear contribution of about $1~\mathrm{pAV^{-1}}$, associated with spurious leakage paths in our setup, was subtracted from the $\Ig$ measurements shown in the Main Text, as described in the Supplementary Information. To avoid potential contributions from displacement currents or hysteresis, $\Ig$ values were recorded by sweeping $\Vg$ from $0~\mathrm{V}$ towards either positive or negative voltages and waiting times in excess of $30~\mathrm{s}$ were allowed. Plots as that of Figs.~\ref{fig:1}(c) and (d) were then obtained by merging two data sets at $\Vg=0~\mathrm{V}$. In case a second gate was present and left grounded, as for the device in Fig.~\ref{fig:1}(a), it was verified that most of the gate current was flowing from the energized gate to the nanowire and not to the grounded gate. 

Measurements in Fig.~\ref{fig:5} were performed via resistive bias-Ts ($50~\mathrm{k\Omega}$ resistors and $22~\mathrm{nF}$ capacitors) mounted on the chip carrier, enabling simultaneous application of DC and AC signals. AC signals were applied via superconducting, non-attenuated coaxial lines. The fast changing gate voltage was applied via an arbitrary waveform generator, while the transmission through the nanowire was measured with a Zurich Instruments Ultra High Frequency Lock-in, operated at a base frequency of $250~\mathrm{MHz}$. An AC signal $V_{\mathrm{in}}$ was applied to the nanowire via a $-80~\mathrm{dB}$ attenuator mounted at the input line of the fridge. The transmitted voltage $V_{\mathrm{out}}$ was measured via the $50~\mathrm{\Omega}$ input port of the lock-in amplifier. The signal $V_{\mathrm{in}}$ had sufficiently low amplitude ($\leq50~\mathrm{\mu V}$) that the highest current flowing in the device was smaller than $\Ir$. The voltage $V_{\mathrm{out}}$ was demodulated by the lock-in amplifier and sent to an oscilloscope with $200~\mathrm{MHz}$ bandwidth. The lock-in demodulator was set to the shortest possible time constant ($30~\mathrm{ns}$), corresponding to a measurement bandwidth of $14~\mathrm{MHz}$. A diagram of the measurement setup is shown in the Supplementary Information.

\bibliography{Bibliography}

\newpage

\section{Supplementary Information}
\setcounter{figure}{0}
\setcounter{equation}{0}
\setcounter{table}{0}
\renewcommand{\thefigure}{S.\arabic{figure}}
\renewcommand{\theequation}{S.\arabic{equation}}

\subsection{Processing of the gate current}
\begin{figure}[b!]
\includegraphics[width=\columnwidth]{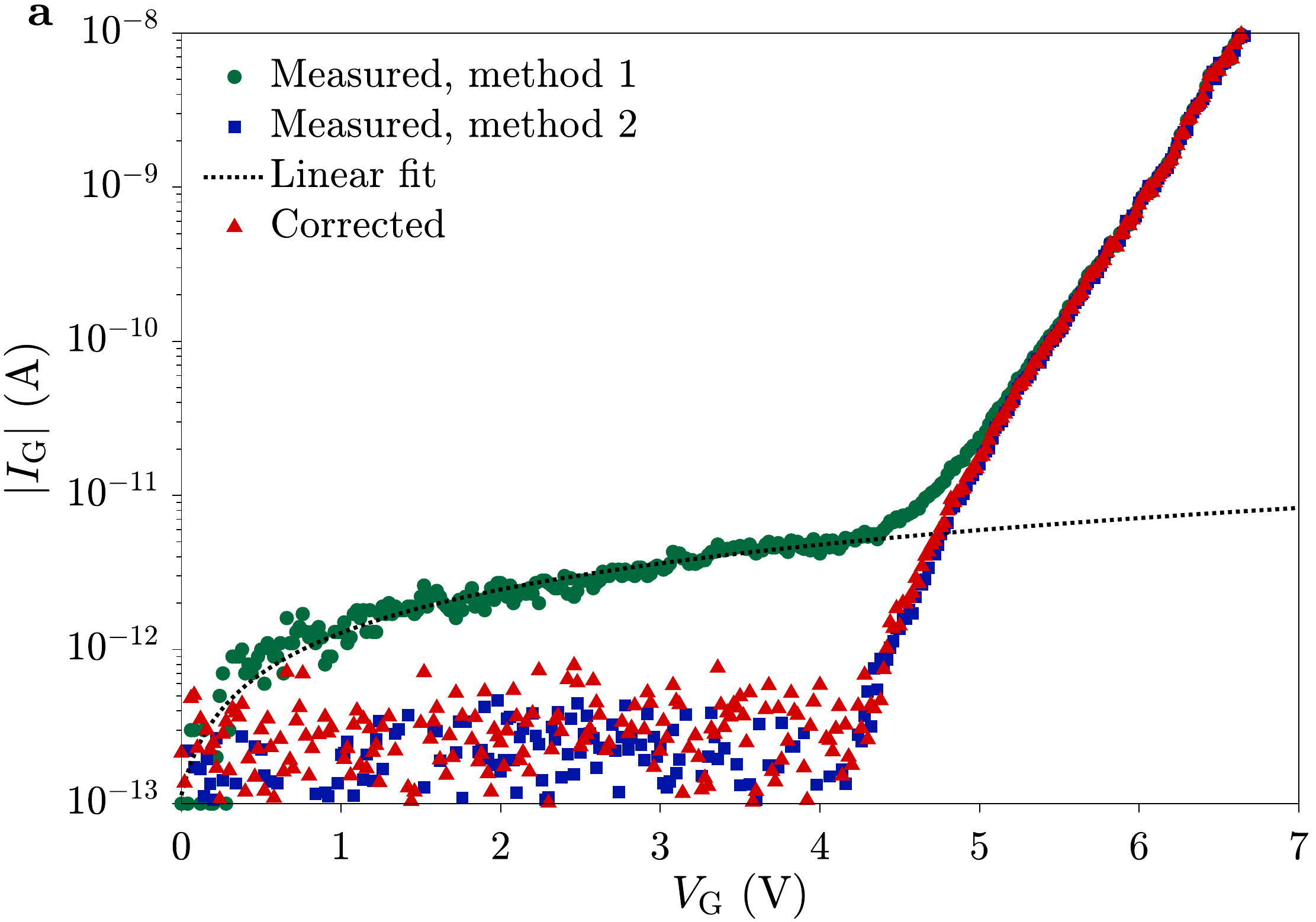}
\caption{
\textbf{Processing of the gate current.} Measurement and processing techniques for the gate current $\Ig$ as a function of gate voltage $\Vg$. Green dots are measured by Method~1 (see text), blue squares by Method~2 (see text). The dotted black line is a linear fit to the green dots for $Vg\leq3.5~\mathrm{V}$. Red triangles are obtained by subtracting the dotted line from the green dots.}
\label{fig:S1}
\end{figure}

Measurements of the gate current $\Ig$, flowing between gate and nanowire channel were obtained with a Keysight B2902A source-measure unit and processed by the numerical technique described below. The source-measure unit applied a voltage $\Vg$ to the gate contact and read the current $\Ig$ flowing into the setup. An example of an $\Ig$ vs. $\Vg$ curve measured with this technique, referred to as Method~1, is shown in Fig.~\ref{fig:S1} (green dots). When measuring with Method~1, we describe the DC current flowing into the electrical setup as sum of two components: the current that actually flows into the gate electrode and reaches the nanowire, and the current that is lost before reaching the gate by spurious leakage paths present in the cryostat. The first current component is expected to depend exponentially on $\Vg$, the second was found to be approximately linear. That is, for small gate voltages, the current flowing to spurious paths and not reaching the gate voltage is dominating. The following numerical procedure was applied to extract the current contribution that actually reaches the nanowire. First, a current offset related to the measurement device is subtracted from the data (typically within $\pm5~\mathrm{pA}$), so that $\Ig=0$ for $\Vg=0$. Second, a line is fit to the experimental $\Ig(\Vg)$ curve for small values of $\Vg$ (dotted black line in Fig.~\ref{fig:S1}). Third, the obtained line is subtracted from the experimental curve in the entire $\Vg$ range, resulting in the red triangles in Fig.~\ref{fig:S1}.
This procedure typically results in gate currents which are within the noise level of our setup at low $\Vg$, and then increase exponentially at large $\Vg$. The linear component that we subtracted corresponds to a resistance of about $1~\mathrm{T\Omega}$. Testing different parts of our setup individually showed that this spurious resistance is predominantly associated with the low frequency twisted pair wires that bring the signal from room temperature to the mixing chamber stage. To verify the validity of this numerical procedure, we measured the current with a second technique, referred to as Method~2. In Method~2, all contacts to the nanowire channel are left floating except for one, which is grounded via a low impedance IV converter (Basel Physics SP 983, with feedback resistance set to $1~\mathrm{G\Omega}$). The gate current $\Ig$ injected into the nanowire channel flows into the IV converter and gives rise to voltage output of $\Ig\times1~\mathrm{G\Omega}$. The current $\Ig$ simultaneously measured with Method~2 is also shown in Fig.~\ref{fig:S1} (blue squares) and is essentially identical to that processed with the numerical technique described above. Adopting Method~2 throughout this work would however not be possible, as the large source-drain currents needed to reach the critical current would result in overloading of the IV converter.

\subsection{Temperature and field dependence of the gate current}
\begin{figure}[t!]
\includegraphics[width=\columnwidth]{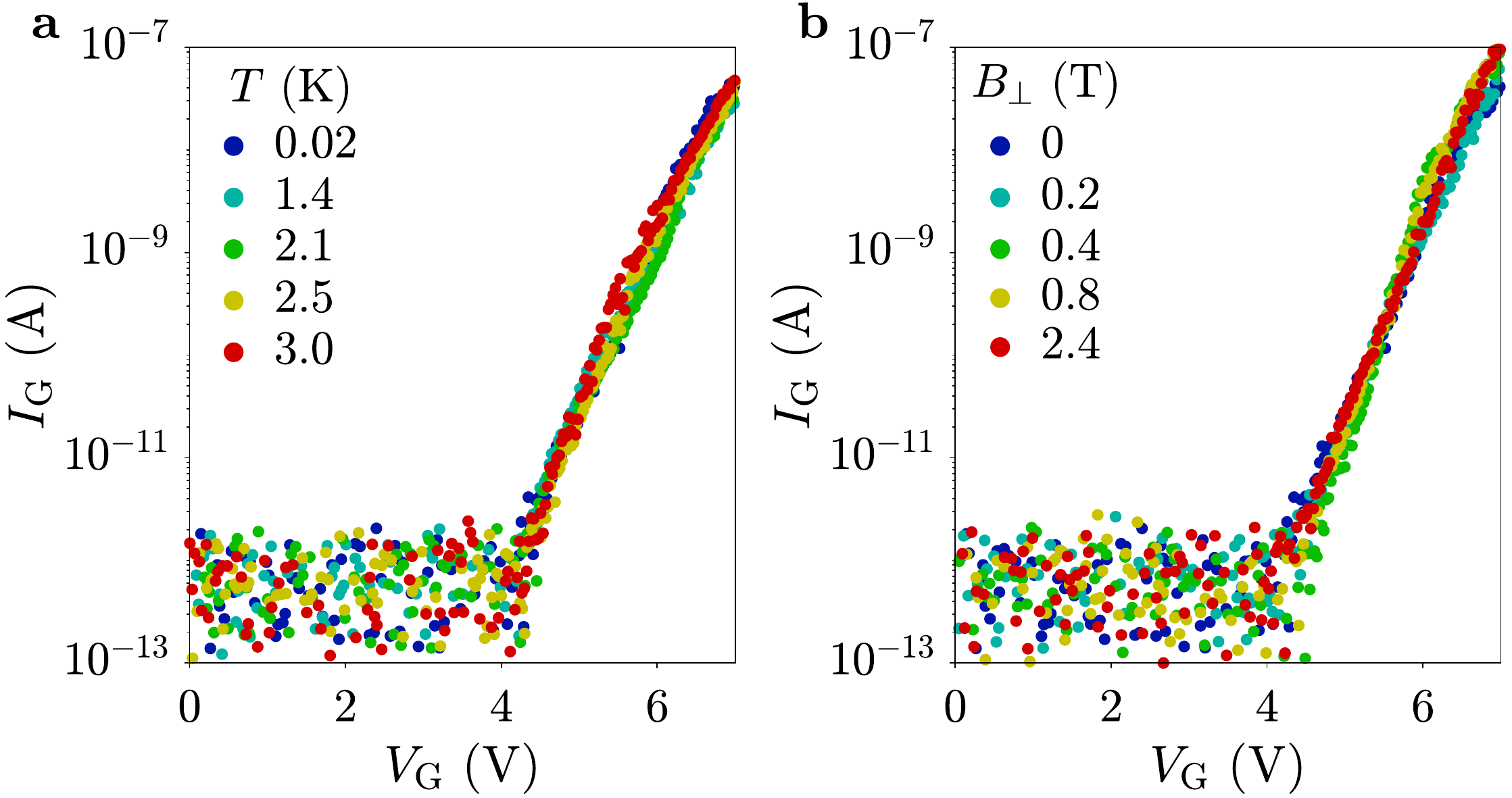}
\caption{
\textbf{Temperature and magnetic field dependence of the gate current.} \textbf{a} Gate current $\Ic$ as a function of gate voltage $\Vg$ simultaneously measured as in Fig.~2(a) of the Main Text. \textbf{B} Gate current $\Ic$ as a function of gate voltage $\Vg$ simultaneously measured as in Fig.~2(b) of the Main Text.}
\label{fig:S2}
\end{figure}

Figure~2 of the Main Text shows the critical current $\Ic$ as a function of gate voltage $\Vg$ for the device of Fig.~1 of the Main Text measured at various temperatures and fields. In Fig.~\ref{fig:S2}(a) and (b) we show the simultaneously measured gate currents $\Ig$. Data indicate that $\Ig$ is unaffected by both temperatures and magnetic fields.

\subsection{Time-resolved measurements}
\begin{figure*}[th!]
\includegraphics[width=2\columnwidth]{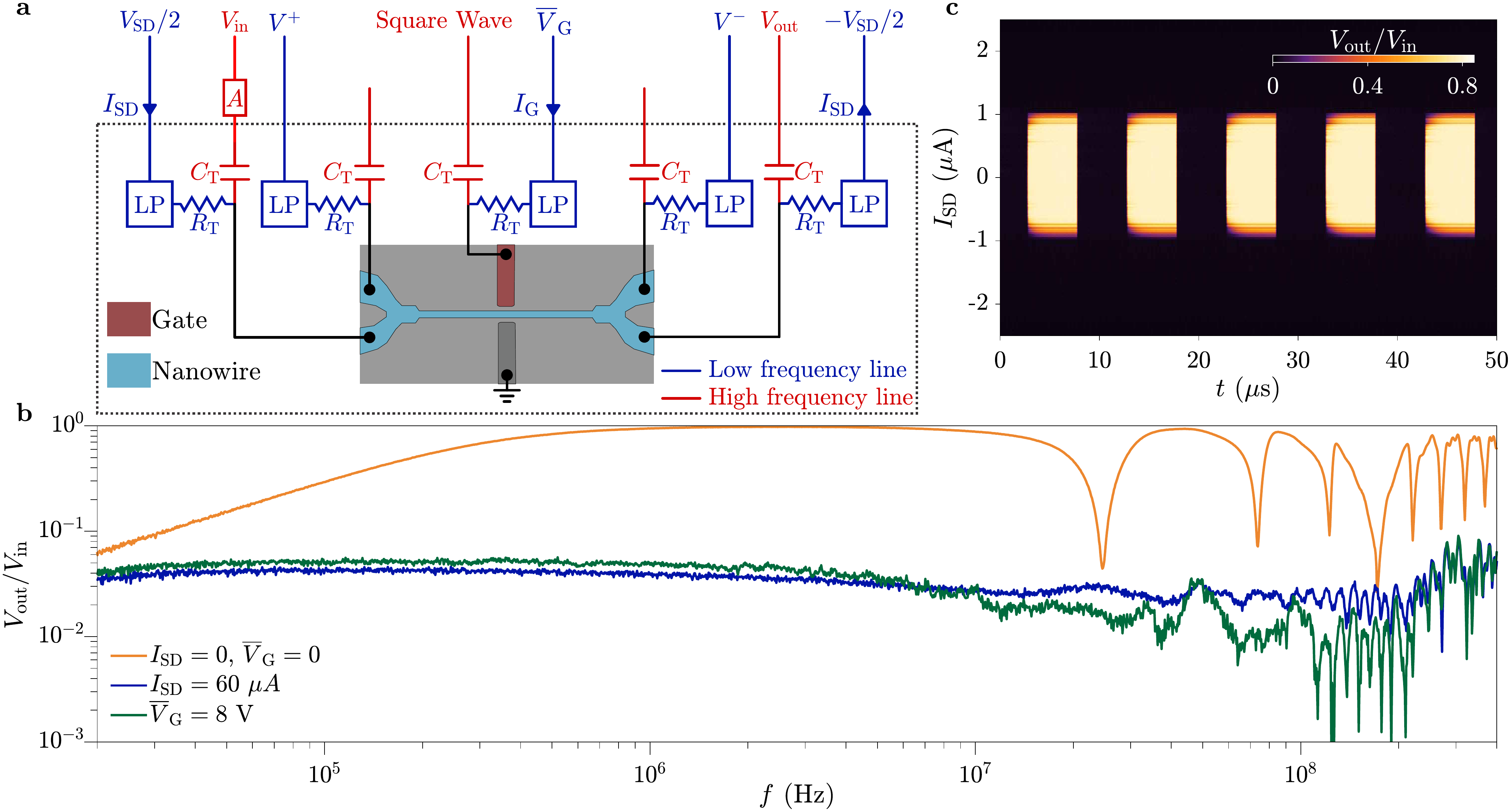}
\caption{\textbf{High frequency measurements.} \textbf{a} Schematic measurement setup for the high frequency experiment presented in Fig.~5 of the Main Text. The dashed box divides the $10~\mathrm{mK}$ (inside) and room temperature (outside) apparatus. Lines for low and high frequency signals are indicated in blue and red, respectively. Low temperature bias-Ts with components $R_{\mathrm{T}}$ and $C_{\mathrm{T}}$ were connected to the four leads of the wire and a side gate. Low frequency cables were filtered by low-pass (LP) filters. Symmetric application of low frequency voltage $\pm\Vsd/2$ results in the flow of a current $\Isd$. Fast signals $V_{\mathrm{in}}$ and $V_{\mathrm{out}}$ are applied and recorded, respectively, thorough $50~\mathrm{\Omega}$ ports. \textbf{b} Frequency dependent $V_{\mathrm{out}}/V_{\mathrm{in}}$ signal measured with the nanowire in the superconducting state (orange), in the normal state as a consequence of a large source-drain current (blue), and in the normal state as a consequence of a large gate voltage (green). \textbf{c} Switching operation as a function of the DC source-drain current $\Isd$. Self-resetting was possible for $\Isd$ values smaller than the nanowire retrapping current.}
\label{fig:S3}
\end{figure*}

\subsection{Measurement of a device with large gates}
\begin{figure*}[th!]
\includegraphics[width=2\columnwidth]{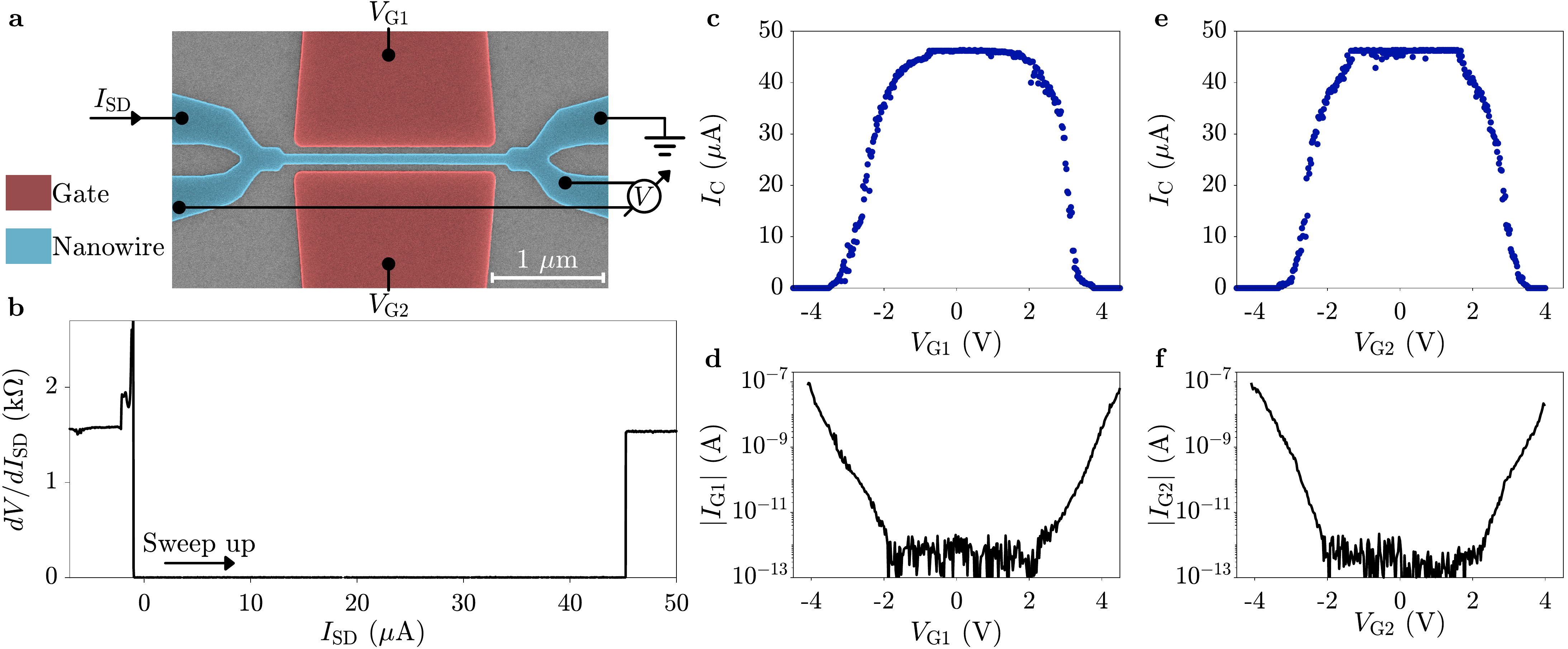}
\caption{\textbf{Measurements of a device with wide side gates.}
\textbf{a} False-color scanning electron micrograph of the device under study and schematics of the measurement setup. The nanowire has been colored blue and the side gates red. \textbf{b} Nanowire resistance $dV/d\Isd$ as a function of source-drain bias current $\Isd$. The measurement highlights the critical current $\Ic=47~\mathrm{\mu A}$ and the retrapping current $\Ir=1~\mathrm{\mu A}$. \textbf{c} Critical current $\Ic$ as a function of gate voltage $V_{\mathrm{G1}}$. \textbf{d} Gate current $I_{\mathrm{G1}}$ as a function of gate voltage $V_{\mathrm{G1}}$. \textbf{e} Critical current $\Ic$ as a function of gate voltage $V_{\mathrm{G2}}$. \textbf{f} Gate current $I_{\mathrm{G2}}$ as a function of gate voltage $V_{\mathrm{G2}}$.}
\label{fig:S4}
\end{figure*}

Figure~5 of the Main Text shows time-resolved measurements, indicating fast switching operation as a function of $\Vg$. A schematic of the electrical setup used for those measurements is shown in Fig.~\ref{fig:S3}(a), which allows for the simultaneous application of low frequency and high frequency signals (colored blue and red, respectively). The device was mounted on a sample holder with resistive bias-Ts (resistance $R_{\mathrm{T}}=50~\mathrm{k\Omega}$ and capacitance $C_{\mathrm{T}}=22~\mathrm{nF}$). The low frequency lines (resistive twisted pairs) passed through RC low pass filters on the sample holder (LP in Fig.~\ref{fig:S3}(a)) and additional RC filters and high frequency pi-filters at the mixing chamber level (not shown), resulting in an additional line resistance $R_{\mathrm{L}}=2.5~\mathrm{k\Omega}$. 
A voltage bias $\Vsd$, symmetrically applied between two low frequency inputs, resulted in a source-drain current of approximately $\Isd=\Vsd/(2R_{\mathrm{T}}+2R_{\mathrm{L}})$. Application of a symmetric bias ensured the nanowire potential was constant with respect to the gate potential as $\Isd$ varied. Low frequency voltage signals $V^+$ and $V^-$ were used to calculate the nanowire four terminal resistance as $R=(V^+-V^-)/\Isd$.

Using the low temperature bias-Ts, a square wave signal was superimposed to the low frequency gate voltage $\oVg$. The device transmission was measured via a lock-in amplifier (Zurich Instruments UHFLI, with input and output set to $50~\mathrm{\Omega}$ impedance.) by applying a voltage $V_{\mathrm{in}}$ through a $-80~\mathrm{dB}$ attenuator ($A$) and recording the resulting voltage $V_{\mathrm{out}}$. The ratio $V_{\mathrm{out}}/V_{\mathrm{in}}$ is shown in Fig.~\ref{fig:S3}(b) as a function of frequency $f$ for three situations. In orange is the situation where the wire was superconducting, meaning $\Isd=0$ and $\Vg=0$. In blue is the situation in which the wire was turned normal by means of a DC current $\Isd=60\mathrm{\mu A}$, larger than the nanowire critical current $\Ic=50~\mathrm{\mu A}$. In green is the situation in which the nanowire was turned normal by the application of a DC gate voltage $\oVg=8\mathrm{V}$. As expected, for sufficiently high frequency the device transmission in the superconducting state approaches unity. Deviations however occur at specific frequencies, presumably due to the fact that the device was not designed to operate at high frequencies. Measurements shown in Fig.~\ref{fig:S3}(c) and in Fig.~5 of the Main Text were performed at a frequency of $250~\mathrm{MHz}$.
Our superconducting switch has the remarkable property to operate without the need of a DC current $\Isd$ flowing in it. Measurements shown in Fig.~5 of the Main Text were obtained with $\Isd=0$, where latching mode is not required. As expected, similar behavior was obtained for $|\Isd|$ smaller than the retrapping current $\Ir$ ($\Ir=1.1~\mathrm{\mu A}$ in the present device). Figure~\ref{fig:S3}(c) shows switching operation as a function of $\Isd$. For $|\Isd|<1~\mathrm{\mu A}$ clear and fast switching operation was obtained, without the need of self-resetting the device at every gate cycle. On the contrary, for $|\Isd|>\Ir$ no switching was observed.

Data shown in the Main Text was obtained on nanowires where gates were relatively narrow (of the order of $100~\mathrm{nm}$ or less) and terminated with a sharp tip. In Fig.~\ref{fig:S4} we present measurements obtained on a nanowire as that of Fig.~1(a) of the Main Text, but with $2~\mathrm{\mu m}$ wide gates. A false-colored scanning electron micrograph of the device is shown in Fig.~\ref{fig:S4}(a), together with the measurement setup. The nanowire is colored in blue and the two gates in red. The gates are separately operated with gate voltages $V_{\mathrm{G1}}$ and $V_{\mathrm{G2}}$, respectively. The response to a source-drain current was characterized in Fig.~\ref{fig:S4}(b) by sweeping $\Isd$ up from the resistive state. As for the wire in Fig.~1(a) of the Main Text, the critical current was $\Ic=47~\mathrm{\mu A}$ and the retrapping current $\Ir=1~\mathrm{\mu A}$. The source-drain critical current $\Ic$ as a function of gate voltage $V_{\mathrm{G1}}$ is shown in Fig.~\ref{fig:S4}(c), with the corresponding gate current $I_{\mathrm{G1}}$ shown in Fig.~\ref{fig:S4}(d). Equivalent measurements performed as a function of gate voltage $V_{\mathrm{G2}}$ are shown in Figs.~\ref{fig:S4}(e) and (f). For both $V_{\mathrm{G1}}$ and $V_{\mathrm{G2}}$, full suppression of $\Ic$ was reached at about $\pm4~\mathrm{V}$, corresponding to gate currents of about $\pm1~\mathrm{nA}$.

\end{document}